\documentclass{moriond}

\def\be{\begin{equation}}
\def\ee{\end{equation}}
\def\bea{\begin{eqnarray}}
\def\eea{\end{eqnarray}}

\newcommand{\p}{\partial}
\def\nn{\nonumber}
\usepackage{amssymb}

\newcommand{\Photo}{}

\begin{document}
\vspace*{4cm}
\title{DARK MATTER RELIC DENSITY REVISITED: THE CASE FOR EARLY KINETIC DECOUPLING}

\author{T. BINDER$^1$, T. BRINGMANN$^2$, M. GUSTAFSSON$^1$,  A. HRYCZUK$^2$\,\footnote{Based on a presentation given at 53$^{\rm{rd}}$ Rencontres de Moriond EW 2018.}}

\address{$^1$Institute for Theoretical Physics, Georg-August University G\"ottingen, Friedrich-Hund-Platz 1, G\"ottingen, D-37077 Germany \\
$^2$Department of Physics, University of Oslo, Box 1048 NO-0316 Oslo, Norway}

\maketitle\abstracts{Kinetic decoupling of dark matter typically happens much later than chemical freeze-out. In fact, local thermal equilibrium is an important assumption for the usual relic density calculations based on solving the Boltzmann equation (for its 0-th moment) describing the dark matter number density. But is this assumption always justified? Here we address this question and discuss the consequences of more accurate treatments. The first treatment is relying on the inclusion of higher moments of the Boltzmann equation and the second on solving the evolution of the phase-space distribution function fully numerically. For illustration, these methods are applied to the Scalar Singlet model, often referred to as the simplest WIMP DM possibility from a model-building perspective. It is explicitly shown that even in this simple model the prediction for the dark matter abundance can be affected by as much as one order of magnitude.
}

\section{Introduction}
\label{sec:intro}

The thermal production through the freeze-out mechanism\,\cite{Gondolo:1990dk} constitutes one of 
the most natural and attractive options to produce the present abundance of dark matter (DM) particles. 
The main assumption entering the standard formalism describing such a process is that, during the time of freeze-out, 
DM is still kept in {\it local} thermal equilibrium (LTE) with the heat bath.

This talk is based on a work\,\cite{Binder:2017rgn} pointing out that exceptions to this standard picture exist, 
where kinetic decoupling happens very early and it cannot be ignored during the freeze-out process. Both 
semi-analytical and fully numerical methods were developed to solve the Boltzmann equation and to compute the DM relic abundance 
in such circumstances.

As an illustration we study in detail the Scalar Singlet model\,\cite{Silveira:1985rk}, for which 
we find an effect on the DM relic density as large as an order of magnitude. The presented methods 
are, however, of much larger generality and can be applied to other scenarios as well. In particular, when studying non-minimal DM models, 
with more than one state in the dark sector, the assumption of LTE is not always well motivated. Both presented methods can be directly used also in such cases, even when the standard treatment is not applicable. 

\section{Thermal production of dark matter}
\label{sec:BEgeneral}

The evolution of the DM phase-space density $f_\chi(t,\mathbf{p})$ is governed by the Boltzmann
equation which in a Friedmann-Robertson-Walker Universe is given by
\be
  \label{diff_boltzmann}
  E\left(\partial_t-H\mathbf{p}\cdot\nabla_\mathbf{p}\right)f_\chi=C[f_\chi]\,,
\ee
where $H$ is the Hubble parameter and the collision term $C[f_\chi]$
describes all interactions between SM particles $f$ and the DM. We are interested, to leading order, in two-body processes for
DM annihilation and elastic scattering, $C=C_{\rm ann}+C_{\rm el}$. Assuming $CP$ invariance 
\bea
  \label{Cann_simp}
  C_\mathrm{ann}= g_{\chi} E\int\frac{d^3\tilde p}{(2\pi)^3} \,v \sigma_{\bar\chi\chi\rightarrow \bar f f}\left[
f_{\chi,{\rm eq}}(E)f_{\chi, {\rm eq}}(\tilde E)-f_\chi(E)f_\chi(\tilde E)
\right]\,, 
\eea
where $v=v_{\rm M\o l}\equiv ({E \tilde E})^{-1}[{(p \cdot \tilde p)^2-m_\chi^4}]^{1/2}$ is the M\o ller velocity. The scattering term is more involved, but in
the non-relativistic limit and assuming that the momentum exchanged in the scattering process is much smaller than the DM mass one finds\,\cite{Binder:2017rgn,Bringmann:2006mu,Binder:2016pnr,Gondolo:2012vh}:
\be
C_\mathrm{el}\simeq 
\frac{E}{2} \gamma(T)
{\Bigg [}
T E \partial_p^2 + \left( p + 2 T  \frac{E}{p} + T \frac{p}{E} \right) \partial_p + 3
{\Bigg ]}f_{\chi}
\label{Csemirel}
\ee
where the momentum exchange rate is given by
\be
  \label{cTdef}
  \gamma(T) =  \frac{1}{3 \pi^2g_\chi m_\chi} \int d\omega\,g^\pm
  \partial_\omega\left( k^4 
  \sigma_T(k)\right),
\ee
and $\sigma_T=\int d\Omega (1-\cos\theta)d\sigma/d\Omega$ is 
the usual transfer cross section for elastic scattering.

\subsection{Coupled Boltzmann equations}
\label{sec:cBE}

The main assumption that enters in the standard treatment\,\cite{Gondolo:1990dk} is the requirement that during chemical freeze-out LTE with the heat bath is maintained. This allows to introduce an Ansatz $f_\chi=A(T)f_{\chi,{\rm eq}}$, where before chemical freeze-out $A(T)=1$. As it is explicitely shown below this assumption is however, even for a standard WIMP, not always justified. In that case in principle one
has to numerically solve the full Boltzmann equation in phase-space, Eq.~(\ref{diff_boltzmann}). However, it sometimes suffices
to take into account the 2nd moment of Eq.~(\ref{diff_boltzmann}), instead of only the 0th moment as in the standard treatment. This leads to a 
manageable {\it coupled} system of differential equations (cBEs).

In analogy to {$Y \equiv n_\chi/s$ for the 0th moment of $f_\chi$, we define 
dimensionless version of the {\it second} moment of $f_\chi$,
\be
\label{ydef}
 y\equiv \frac{m_\chi}{3 s^{2/3}}
\left\langle \frac{\mathbf{p}^2}{E} \right\rangle
=
\frac{m_\chi}{3 s^{2/3}}
\frac{g_\chi}{n_\chi}\int \frac{d^3p}{(2\pi)^3}\,\frac{\mathbf{p}^2}{E} f_\chi(\mathbf{p})\,.
\ee
If DM particles follow the thermal distribution, e.g. by sufficiently strong self-scattering, they have a {\it temperature} $T_\chi = y s^{2/3}/m_\chi$.
In general, for non-thermal distributions, one can read the above equation as an {\it definition} of DM 'temperature', in terms of the 2nd moment of $f_\chi$.

Integrating Eq.~(\ref{diff_boltzmann}) over $g_\chi\int d^3p/(2\pi)^3/E$ and $g_\chi\int d^3p/(2\pi)^3\mathbf{p}^2/E^2$,
respectively, and plugging in $C=C_{\rm ann}+C_{\rm el}$ given in Eqs.~(\ref{Cann_simp},\ref{Csemirel}) we arrive at:
\bea
\frac{Y'}{Y} &=& \frac{s Y}{x \tilde H}\left[
\frac{Y_{\rm eq}^2}{Y^2} \left\langle \sigma v\right\rangle- \left\langle \sigma v\right\rangle_{\rm neq}
\right], \label{Yfinalfinal}\\
\frac{y'}{y} &=&   \frac{\gamma(T)}{x\tilde H}\left[\frac{y_{{\rm eq}}}{y} -1\right]
+\frac{sY}{x\tilde H}\left[
\left\langle \sigma v\right\rangle_{\rm neq}-\left\langle \sigma v\right\rangle_{2,{\rm neq}}
\right] \label{yfinalfinal}\\ 
&&+\frac{sY}{x\tilde H}\frac{Y_{\rm eq}^2}{Y^2}\left[
\frac{y_{{\rm eq}}}{y}\left\langle \sigma v\right\rangle_{2}-\left\langle \sigma v\right\rangle
\right]
+\frac{H}{x\tilde H} \frac{\langle p^4/E^3 \rangle}{3T_\chi}\,. \nonumber
\eea
Here, in addition to usual $\langle \sigma v\rangle$, we also made use of temperature-weighted thermal average:
\bea
\label{therm_av2}
 \left\langle \sigma v\right\rangle_{2}
 &\equiv& 
 \frac{g_\chi^2}{T n_{\chi,{\rm eq}}^2} \int 
 \frac{d^3p\, d^3\tilde p}{(2\pi)^6}
 \frac{p^{2}}{3E} 
 \sigma v_{\bar\chi\chi\rightarrow \bar f f} f_{\chi,{\rm eq}}({\mathbf{p}}) f_{\chi,{\rm eq}}(\tilde{\mathbf{p}}).
\label{therm_av2eq}
\eea
The `non-equilibrium average' $\left\langle \sigma v\right\rangle_{2,{\rm neq}}$ is understood to be defined as 
in Eq.~(\ref{therm_av2}), but for an arbitrary  $n_{\chi}$, $f_\chi(\mathbf{p})$ and hence also
$1/T\to1/T_\chi$ in the normalization prefactor. 

The set of Eqns.~(\ref{Yfinalfinal}, \ref{yfinalfinal}) includes even higher moment of $f_\chi$, 
and hence does not close w.r.t.~the variables $Y$ and $y$. We need additional input
to determine the quantities $\left\langle \sigma v\right\rangle_{{\rm neq}}$, 
$\left\langle \sigma v\right\rangle_{2,{\rm neq}}$ and $\langle p^4/E^3 \rangle$
in terms of only $y$ and $Y$. We will make the following ansatz for these quantities:
\bea
  \left\langle \sigma v\right\rangle_{{\rm neq}} &=&  \left.\left\langle \sigma v\right\rangle\right|_{T=y s^{2/3}/m_\chi}, 
  \label{svansatz}\qquad
\left\langle \sigma v\right\rangle_{2,{\rm neq}} = \left.\left\langle \sigma v\right\rangle_{2}\right|_{T=y s^{2/3}/m_\chi},
\label{sv2ansatz}\\
\langle p^4/E^3 \rangle &=&
\left[\frac{g_\chi}{2\pi^2 n_{\chi,\rm{eq}}(T)}\int dp\frac{p^6}{E^3} e^{-\frac{E}{T}}\right]_{T=y s^{2/3}/m_\chi}
\label{p4E3ansatz}\,.
\eea
These expressions would result from an equilibrium DM phase-space distribution but at a temperature different from that of the heat bath. 


\subsection{The full phase-space density evolution}
\label{sec:fBE}

The second method applicable even if LTE is not maintained around freeze-out is to solve the Boltzmann Eq.~(\ref{diff_boltzmann}) at the full phase-space density level. We start by rewriting Eq.~(\ref{diff_boltzmann}) in two dimensionless coordinates
$
x(t,p)\equiv m_\chi/T   \ \rm{and} \ q(t,p) \equiv p/T,
$
where $q$ is now the `momentum' coordinate that depends on both $p$ and $t$. Such new coordinates absorb exclusively the change of the DM momentum and density 
 due to the Hubble expansion. In these variables Eq.~(\ref{diff_boltzmann}) becomes
\bea
\p_x f_\chi(x,q)
 &=& \frac{m_\chi^3}{\tilde{H} x^4}\frac{g_{\bar \chi} }{2\pi^2}\int{\!d\tilde q\;\tilde q^2}\;  \frac{1}{2}\!\int{\!d\!\cos{\theta}\,}  \; v \sigma_{\bar\chi\chi\rightarrow \bar f f}  \left[ f_{\chi,{\rm eq}}(q)f_{\chi,{\rm eq}}(\tilde q)-f_\chi(q)f_\chi(\tilde q) \right] \nn\\
&+& \frac{\gamma(x)}{2 \tilde{H} x} \left[x_q \partial^2_q + \left(q+\frac{2 x_q}{ q} + \frac{q}{x_q}\right)\partial_q+3\right]f_\chi +  \tilde{g}   \frac{q}{x}  \partial_q f_\chi,
	\label{eq:BEps}
\eea
where $\theta$ is the angle between $\bf{q}$ and $\bf{\tilde q}$, and $x_q \equiv \sqrt{x^2 +q^2}$.\\[-2ex]
 
We then discretize the momentum variable $q$ into $q_i$ with $i\in \{1,2,\ldots,N\}$ what allows us to rewrite the above  \emph{integro partial differential equation}  into a set of $N$ coupled ODEs:
\bea
\frac{d}{dx} f_i &=& \frac{m_\chi^3}{\tilde{H} x^4}\frac{g_{\bar \chi} }{2\pi^2}
\sum_{j=1}^{N-1} 
\frac{\Delta \tilde q_j}{2} 
\Big[
   {\tilde q_j^2 } \,\langle v_{\rm M\o l} \sigma_{\bar\chi\chi\rightarrow \bar f f} \rangle_{i, j}^{\theta}
   \left( f^{\rm eq}_i f^{\rm eq}_j \!-\! f_if_j \right) \nn\\
&+&     {\tilde q_{j+1}^2 } \,\langle v_{\rm M\o l} \sigma_{\bar\chi\chi\rightarrow \bar f f} \rangle_{i, {j+1}}^{\theta}
 \left( f^{\rm eq}_i f^{\rm eq}_{j+1} \!-\! f_if_{j+1} \right)
\Big]
  \nn\\
&+&  \frac{\gamma(x)}{2 \tilde{H} x} \left[x_{q,i} \partial^2_q f_i+ \left(q_i\!+\!\frac{2 x_{q,i}}{q_i} \!+\! \frac{q_i}{x_{q,i}}  \right)\partial_q f_i+3 f_i\right] +    \tilde{g}   \frac{q_i}{x}  \partial_q f_i,
\label{eq:BE_num}
\eea
where $f_i \equiv f_\chi(x,q_i)$, and the derivatives $\p_q f_i$ and $\p^2_q f_i$ are determined numerically 
from several neighboring points to $f_i$.
$\langle v_{\rm M\o l} \sigma_{\bar\chi\chi\rightarrow \bar f f} \rangle_{i, j}^{\theta}$ is the
velocity-weighted cross section averaged over $\theta$ and $\Delta \tilde q_j \equiv  \tilde{q}_{j+1}- \tilde{q}_j$.
We typically used the 
range $q_1 =10^{-6}$ to $q_N=50$ with $\sim 1000$ steps in between. By the use of  the 
ODE15s code in MatLab, and by analytically deriving internally required  Jacobians, we are 
able to efficiently (on the scale of $\sim$min) calculate 
the full phase-space evolution during the freeze-out. The code is general enough to be adapted to any standard single WIMP setup.

\section{Scalar Singlet Dark Matter}
\label{sec:singlet}

The simplest WIMP DM possibility from a model-building perspective is the Scalar Singlet model\,\cite{Silveira:1985rk}.
In it, the only new addition to the Standard Model (SM) is one real gauge-singlet scalar field $S$
which is stabilized by a $\mathbb{Z}_2$ symmetry. The Lagrangian for this model reads
\begin{equation}
{\cal L}_{\rm SZ} = {\cal L}_{\rm SM} + \frac{1}{2} \partial_\mu S \partial^\mu S - \frac{1}{2} \mu_S^2 S^2 - \frac{1}{2} \lambda_S S^2 H^\dagger H - \frac{1}{4!} \lambda_{SS} S^4,
\end{equation}
where here $H$ is the SM Higgs doublet.

Recently, the GAMBIT collaboration presented a global fit of this model taking into account all available experimental constraints\,\cite{Athron:2017kgt}. They find 
the parameter region with the highest profile likelihood to the be one where $m_S \sim m_h/2$, i.e. the DM abundance is mostly set by the resonant annihilation through an almost on-shell Higgs boson. 

In this model the annihilation cross section to SM particles (except $hh$) is given by\,\cite{Cline:2013gha}
\begin{equation}
\sigma v_{\rm CMS} = \frac{2 \lambda_S^2 v_0^2}{\sqrt{s}} \, |D_h(s)|^2 \, \Gamma_{h\to \rm{SM}}(\sqrt{s}), \quad {\rm{where}} \quad |D_h(s)|^2 = \frac{1}{(s-m_h^2)^2+m_h^2 \Gamma_h^2 }
\label{eq:sigmav}
\end{equation}
and $\Gamma_{h\to \rm{SM}}(\sqrt{s})$ is the partial decay width of a would-be SM Higgs boson of mass $\sqrt{s}$.

The elastic scattering processes are $t$-channel Higgs mediated scatterings on all SM fermions. The corresponding squared amplitude takes a simple form,
\be
\left| \mathcal{M}_{Sf\to Sf}\right|^2 = \frac{N_f\lambda_S^2m_f^2}{2}\frac{4m_f^2-t}{(t-m_h^2)^2} \,,
\ee
where $m_f$ is the mass of the SM fermion and the color factor is $N_f= 3(1)$ for quarks (leptons).

The scattering rate is dominated, due to the hierarchical Yukawa structure of the Higgs couplings, 
by the interactions with these fermions that are the heaviest, but at the same time 
still sufficiently abundant in the plasma for a given temperature. Therefore, the precise treatment of heavy 
quarks in the plasma at temperatures around $T\sim \mathcal{O}(1\, \rm{GeV})$ can have a significant impact 
on the scattering rate. To take this into account, we follow the
literature\,\cite{Gondolo:2012vh,Boyanovsky:2006bf} and adopt two extreme scenarios that are bracketing the actual size of 
the scattering term: 'A' - all quarks are unbound and present in the plasma down to $T_c = 154$\,MeV (largest scattering 
scenario) and 'B' - only light $u$, $d$ and $s$ quarks are free and only for temperatures above 
$4T_c \sim 600$ MeV (smallest scattering scenario).

\subsection{Relic density of scalar singlet dark matter}
\label{sec:singlet_rd}

We compute the relic density following both methods described above and compare it to the standard treatment 
adopted in the literature. The results for the relic density and the effect of the proper treatment of the 
kinetic decoupling in the $(m_S,\lambda_S)$ plane are shown in Fig.~\ref{fig:res}. The blue dotted line denotes 
the standard result, as can also be found in the literature. The red solid (dashed) line shows the solution of the 
coupled system of Boltzmann equations (\ref{Yfinalfinal},\ref{yfinalfinal}), for the `B' (`A') scenario for scatterings on quarks.
The black dots give the result for the full numerical solution of the Boltzmann equation in phase-space.

Outside the resonance region, the cBE lead to identical results as
the standard approach, indicating in that case that the assumption 
of LTE during chemical freeze-out thus is well satisfied. On the other hand, close to the resonance region 
we see a large effect, implying that this assumption must be violated. The size of the effect is directly related to the size of the scattering rate and hence to just
exactly how early kinetic decoupling happens -- a smaller rate (as in scenario `B') leads to a larger deviation than the maximal scattering rate (scenario `A'). Let us stress 
that this is an important general message, with implications way beyond the specific model studied.

\begin{figure}[h!]
  \includegraphics[width=0.48\linewidth]{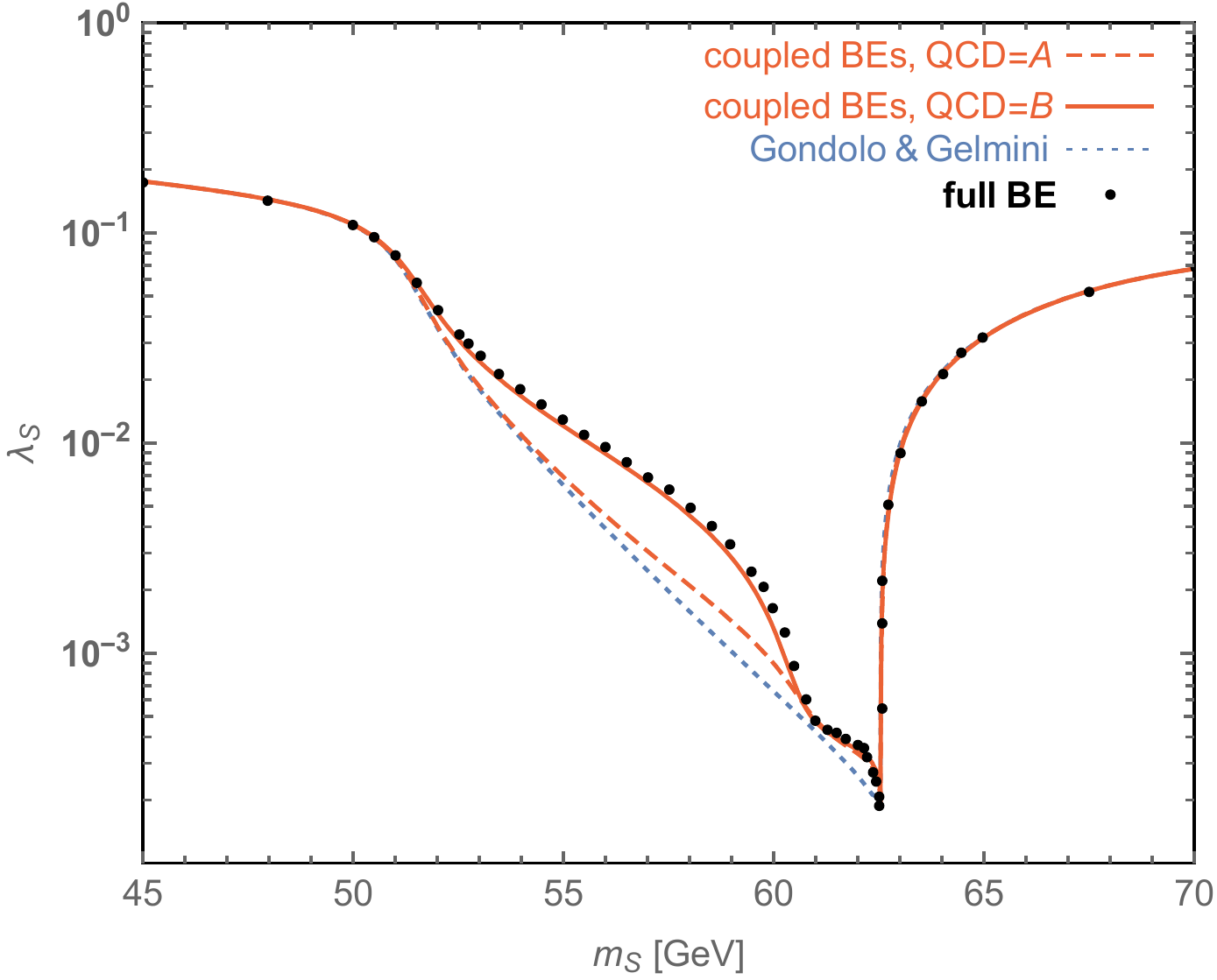}
    \hspace*{0.5cm}
  \includegraphics[width=0.48\linewidth]{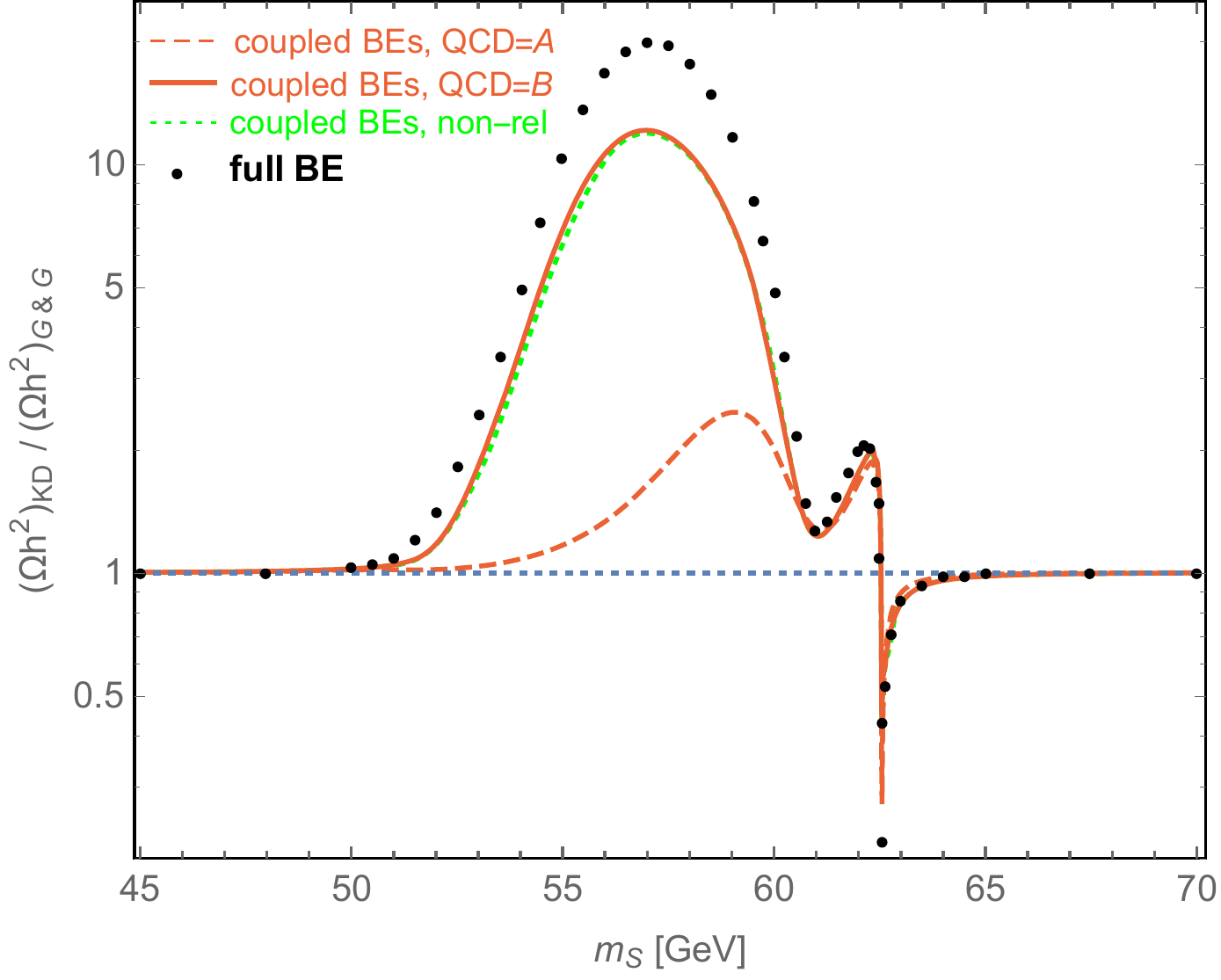}
  \caption{\textit{Left:} the required value of the coupling $\lambda_S$ as a function of the $S$ mass
   giving a thermal relic density of $\Omega h^2=0.1188$.  The blue dotted line shows the standard 
   result, based on the assumption of LTE during freeze-out. The solid and dashed red lines, respectively, give the result of solving instead the coupled system of Boltzmann equations (\ref{Yfinalfinal}) and (\ref{yfinalfinal}) for the maximal (`A') and minimal (`B') quark scattering scenarios.  The fully numerical result of the Boltzmann equation assumes minimal quark scattering and is shown as black dots (`full BE'). \textit{Right:} the effect of the improved treatment of the kinetic decoupling on the results for relic density. Computed for parameter points that would satisfy the relic density in the standard approach (dotted line in the left panel).
  }
    \label{fig:res}
\end{figure}

Moreover, the cBEs (\ref{Yfinalfinal},\ref{yfinalfinal}) provides a qualitatively and often quantitatively very good description for the final DM
abundance, capturing most, if not all, of the effect of the kinetic decoupling. Nevertheless, for high-precision results one needs to actually solve
the full Boltzmann equation in phase-space. This is because, as the full numerical solution reveals, the shape of $f_\chi(p)$ can be quite 
different from the Maxwell-Boltzmann form, which can introduce departure from the assumptions used in the cBEs. 
This can have a seizable impact on the result for the relic density (as for $m_S\sim57$\,GeV) or a very modest one (as for $m_S\sim m_h/2$)
depending on whether or not the shape during chemical freeze-out is affected for momenta that can combine to $\sqrt{s}\sim m_h$.

\begin{figure*}
  \includegraphics[width=0.48\linewidth]{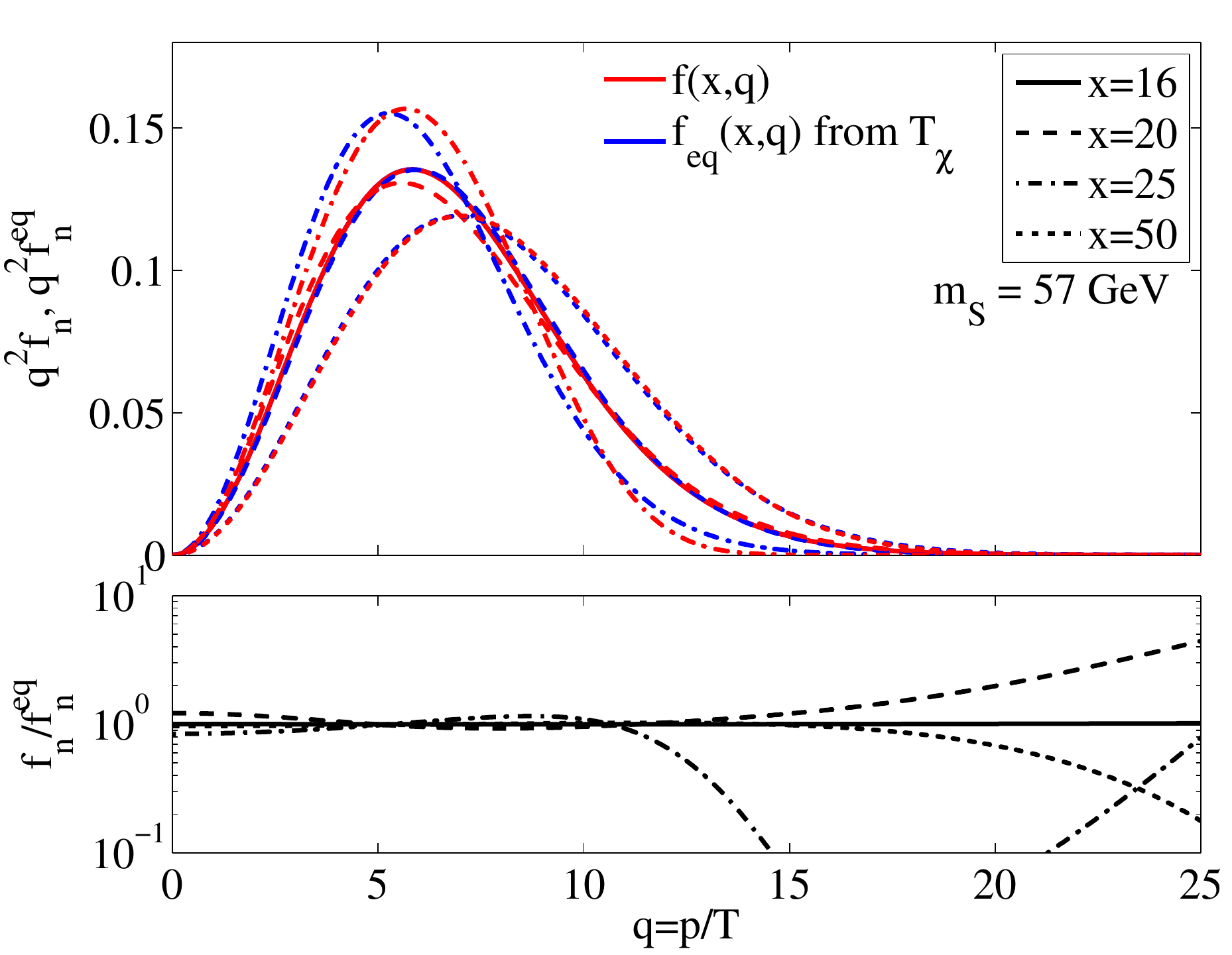}
  \hspace*{0.5cm}
  \includegraphics[trim={1.cm 1.5cm 1.5cm 1.cm}, width=0.48\linewidth]{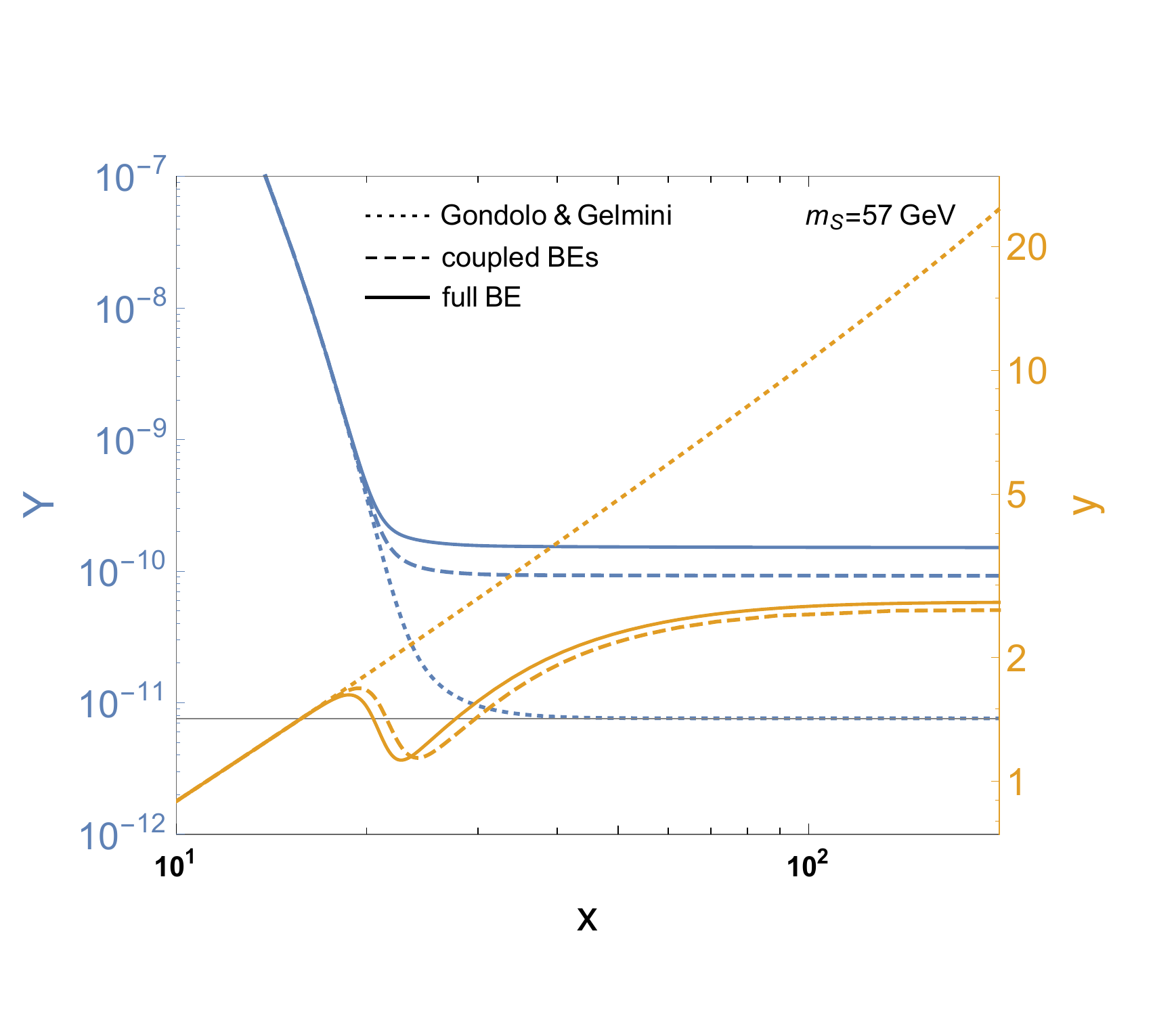}
  \hspace*{0.75cm}
  \caption{Phase space distributions for a scalar singlet DM particle with $m_S=57$\,GeV assuming a Higgs-scalar coupling that leads to the correct relic density in the standard treatment (dotted blue line in left panel of Fig.~\ref{fig:res}). {\it Left:} evolution of unit normalized phase-space distributions $f_n(q)$ from the full numerical solution of the Boltzmann equation (red lines) and equilibrium distributions $f^{\rm{eq}}_n(q)$ at the `temperatures' $T_\chi$ (blue lines), evaluated at four different temperatures $x = m_S/T = $ 16 (solid), 20  (dashed), 25 (dot-dashed) and 50 (dotted). The bottom panel highlights the deviation from the corresponding thermal distribution by plotting $f_n(q) / f_n^{\rm{eq}}(q)$. {\it Right:} the respective evolution of $Y$ (blue) and $y$ (yellow) for the standard case (dotted lines), the approach using cBEs (dashed) and the full numerical result (solid).}
    \label{fig:PSdist}
\end{figure*}

For illustration, in Fig.~\ref{fig:PSdist} we take an example case of $m_S=57$\,GeV and show the full 
phase-space distribution for a few representative values of $x$ as well as the 
corresponding evolution of $Y$ and $y$. This parameter point exhibits a relatively large difference between full solution and cBEs, as visible in 
Fig.~\ref{fig:res}. Fig.~\ref{fig:PSdist} shows that this difference arises because of the dip in the ratio of DM 
phase-space distributions that starts to develop around the freeze-out time, for $x\gtrsim20$. This dip in turn originates due to the resonance enhancing the annihilation in this momentum range for these $x$ values. As seen on the right panel of Fig.~\ref{fig:PSdist}, this 
results in the DM particles falling out of chemical equilibrium earlier, and therefore enhance the value of the thermal relic density}. It is important to stress that the bulk of this effect is well captured by the cBEs (compare the dashed vs. solid lines in the right panel of Fig.~\ref{fig:PSdist}).

\section{Conclusions}
\label{sec:conc}

We have shown that very early kinetic decoupling is something more than just a theoretical possibility. Indeed, we 
demonstrated that departure from kinetic equilibrium can instead happen much earlier, even at the same time as the departure from chemical equilibrium.
Moreover, this can appear even in simple WIMP models and can affect the DM relic density in a very significant way. 
Therefore, the standard way of calculating the thermal relic density needs to be extended, as it rests on the assumption of local thermal equilibrium during freeze-out.
We provide two methods for dealing with this issue, one introducing a coupled system of equations for the DM number density and its
`temperature', and second relying on full numerical solver of the phase-space Boltzmann equation.  The latter approach
has the additional advantage of giving as a result the full $f_\chi(t,\mathbf{p})$, which in particular allows to test the assumption of an equilibrium distribution adopted in the standard treatment.

Let us stress, that both the derived coupled Boltzmann equations and the developed numerical setup are very general, and can be used to perform accurate studies of early kinetic decoupling and thermal relic density for a much larger range of models. Beyond obvious applications to other cases with resonant annihilation (see also\,\cite{Duch:2017nbe}), further examples include Sommerfeld-enhanced 
annihilation \cite{Binder:2017lkj}
, annihilation to DM bound states
, models with large semi-annihilations 
 or scenarios that rely on $3\to2$ or $4\to2$ annihilation processes
.

Finally, the developed numerical code for solving the full Boltzmann equation in the phase-space is going to be publicly released in a separate work.\cite{inprogress} 

\section*{Acknowledgments}

AH is supported by the University of Oslo through the Strategic Dark Matter Initiative (SDI). 
MG and T.Binder have received funding from the European Union's Horizon 2020 research and innovation programme under grant agreement No 690575 and No 674896. T.Binder gratefully acknowledges financial support from the German Science Foundation (DFG RTG 1493).

\end{document}